\documentclass[12pt]{iopart}
\usepackage{iopams}
\newcommand{\Rset}{\mathbb{R}}

\newcommand{\cA}{\mathcal{A}}
\newcommand{\cB}{\mathcal{B}}

\newcommand{\rI}{\mathrm{I}}
\newcommand{\rII}{\mathrm{II}}
\newcommand{\rL}{\mathrm{L}}

\newcommand{\cW}{\mathcal{W}}

\newcommand{\tpi}{\tilde{\pi}}

\newcommand{\tT}{\tilde{T}}
\newcommand{\tA}{\tilde{A}}
\newcommand{\tB}{\tilde{B}}
\newcommand{\tlambda}{\tilde{\lambda}}

\newcommand{\hphi}{{\hat{\phi}}}

\newcommand{\hT}{{\hat{T}}}
\newcommand{\hP}{{\hat{P}}}
\newcommand{\hy}{{\hat{y}}}
\newcommand{\hq}{{\hat{q}}}
\newcommand{\hr}{{\hat{r}}}
\newcommand{\hb}{{\hat{b}}}
\newcommand{\hw}{{\hat{w}}}

\newcommand{\hA}{{\hat{A}}}
\newcommand{\hB}{{\hat{B}}}
\newcommand{\hW}{{\hat{W}}}

\newcommand{\Lag}[2]{L_{#2,#1}}      
\newcommand{\XLagI}[3]{L^{\rI}_{#2,#3,#1}}  
\newcommand{\XLagII}[3]{L^{\rII}_{#2,#3,#1}}  
\newcommand{\cL}{\mathcal{L}}
\newcommand{\cLI}[1]{\cL^{\rI}_{#1}}
\newcommand{\cLII}[1]{\cL^{\rII}_{#1}}

\newtheorem{lem}{Lemma}[section]

\begin{document}

\title[$X_m$-Laguerre polynomials]{Exceptional  orthogonal polynomials and the Darboux transformation.}
\author{D G\'omez-Ullate, N Kamran and R Milson}
\address{ Departamento de F\'isica Te\'orica II, Universidad Complutense de Madrid, 28040 Madrid, Spain}
\address{Department of Mathematics and Statistics, McGill University
Montreal, QC, H3A 2K6, Canada}
\address{Department of Mathematics and Statistics, Dalhousie University, Halifax, NS, B3H 3J5, Canada}
\eads{\mailto{david.gomez-ullate@fis.ucm.es},
\mailto{nkamran@math.mcgill.ca}, \mailto{milson@mathstat.dal.ca}}

\begin{abstract}
  We adapt the notion of the Darboux transformation to the context of
  polynomial Sturm-Liouville problems.  As an application, we
  characterize the recently described $X_m$ Laguerre polynomials in
  terms of an isospectral Darboux transformation.   We also show that
  the shape-invariance of these  new polynomial families is a direct
  consequence of the permutability property of the Darboux-Crum transformation.
\end{abstract}
\maketitle
\section{Introduction}\label{sec:Intro}

The Darboux-Crum transformation is a well-known and powerful technique
in Quantum Mechanics for generating new exactly solvable potentials
from known ones \cite{GKMDarboux1, GKMDarboux2}.  
A situation of particular interest arises when the Darboux
transformation can also be applied to construct new families of
orthogonal polynomials from known ones, since it is often the case
that the bound states of an exactly solvable potential are polynomial
after a change of independent variable and a rescaling of the wave
function by a suitable non-vanishing weight function. Of course, care
has to be exercised in order to characterize those cases in which the
eigenfunctions obtained by this procedure do indeed give rise to
orthogonal polynomial families which will in turn correspond to a
well-defined Sturm-Liouville problem. In particular, once it is known
that the transformed eigenfunctions are polynomial, one still has to
show that they are complete in the underlying weighted $\rL^2$
space. Our purpose in this paper is to apply these ideas to generate
novel families of complete orthogonal polynomial systems that are
solutions of Sturm-Liouville problems, and in particular to show how
the families of generalized Laguerre orthogonal polynomials recently
constructed by Odake, Sasaki \cite{sasaki1,sasaki2,sasaki3,sasaki4},
and Quesne \cite{quesne1,quesne2} fit into the classification
principle for Darboux transformations first introduced in
\cite{GKMDarboux1}. It is worth to stress that these new exceptional polynomial families, although solutions of a 
Sturm-Liouville problem, are outside the Askey-Wilson class \cite{askey}.

To set our results in context, we first recall the foundational theorem of Bochner \cite{Bo} which states
that if an infinite sequence of polynomials $\{P_n(z)\}_{n=0}^\infty$
satisfies a second order eigenvalue equation of the form
\begin{equation}\label{eq:bochner}
  p(x)P_n''(x) + q(x) P_n'(x) + r(x) P_n(x)=\lambda_n P_n(x),\qquad
  n=0,1,2,\dots  
\end{equation}
then $p(x),q(x)$ and $r(x)$ must be polynomials of degree $2,1$ and $0$ respectively.  In addition, if the $\{P_n(x)\}_{n=0}^\infty$
sequence is an orthogonal polynomial system, then it has to be (up to
an affine transformation of $z$) one of the classical orthogonal
polynomial systems of Jacobi, Laguerre or Hermite
\cite{Aczel,Mikolas,Feldmann,Lesky,KL97}.

In a pair of recent papers \cite{GKM09a,GKM09b}, we have shown that
there exist complete orthogonal polynomial systems, defined by
Sturm-Liouville problems, that extend beyond the classical families of
orthogonal polynomials arising from Bochner's classical theorem on the
characterization of Sturm-Liouville polynomial systems. What
distinguishes our hypotheses from those made by Bochner is that the
first eigenpolynomial of the
sequence need not be of degree zero, even though the full set of
eigenfunctions still forms a basis of the weighted $\rL^2$ space.  The
situation we considered in \cite{GKM09b} is that of complete
orthogonal polynomial systems starting in degree one. For this $m=1$
case a full characterization of all Sturm-Liouville polynomial systems
is available thanks to the classification of codimension one
exceptional polynomial subspaces performed in \cite{GKM09a}. The
concept of an exceptional polynomial subspace was introduced in
\cite{GKM3,GKM6}.  For some recent applications of exceptional
orthogonal polynomials see \cite{Midya-Roy,Tanaka}.

In the present paper, we pursue this program in higher codimension by
constructing explicit examples of complete systems of higher
codimension using the Darboux transformation in a systematic
fashion. In particular, we will construct the analogues in codimension
$m$ of the $X_{1}$ Laguerre polynomials.  Our paper is organized as
follows. In Section 2, we define the notion of an $m$-orthogonal
polynomial system arising from modules of higher codimension and we
introduce the notion of polynomial Sturm-Liouville problem,
corresponding to the case in which the eigenfunctions of the
Sturm-Liouville operator are polynomials. In Section 3, we carefully
study the various cases in which the Darboux transformation preserves
the polynomial character of the eigenfunctions (these are precisely
the {\em algebraic Darboux transformations} introduced in
\cite{GKMDarboux1}). We also show the role played by the property of
shape invariance in the various factorizations that give rise to
Darboux-Crum transformations. In Section 4, we apply these results to
the case of the Sturm-Liouville problem defining the Laguerre
polynomials and show precisely how the L1 and L2 families of
codimension $m$ Laguerre polynomials obtained in \cite{sasaki3} fit
into our general classification scheme. In Section 5, we show that
these polynomials satisfy remarkable shape invariance properties that
arise from the intertwining relations obtained between the partner
second-order operators and iterations of the first-order operators
defining the Darboux transformations. We will see that the shape
invariance reported in \cite{sasaki1} \cite{quesne1} follows from the
shape invariance of the initial operator and the formal properties of
the Darboux-Crum transformations.

The perspective taken in this paper is largely that of the formal
calculus of differential operators. Previously, analytic aspects of
the Darboux-Crum method for Sturm-Liouville systems were considered in
\cite{gesztesy}.  Our approach is different in that we focus on
algebraic properties and exact solutions. In a subsequent paper
\cite{GKM10} we shall study the codimension $m$ Laguerre polynomial
system in a functional analytic setting by giving a spectral theoretic
characterization of the codimension $m$ Laguerre polynomial system in
the context of Sturm-Liouville theory.  A detailed analysis of the
asymptotic properties of the zeros of these polynomials will be
given. We shall also further develop some of the key formal properties
of these polynomials that result from the factorization and shape
invariance of their defining operators, namely their orthogonality
properties, Rodrigues-type formulas and generating functions. Finally
we mention that this entire analysis can also be carried out in the
case of Jacobi polynomials.

                                                                                
                                                                                

\section{Preliminaries}
\label{sec:Main}
We will say that a differential operator
\begin{equation}
  \label{eq:Tdef}
  T(y) = p(x) y'' + q(x)y' + r(x)y,\quad y=y(x)
\end{equation}
is exactly solvable by polynomials (PES) if it admits infinitely many
real, polynomial eigenfunctions $y_j(x)$:
\begin{equation}
  \label{eq:Tyjeigenfunctions}
  T(y_j) = \lambda_j y_j,\quad \deg y_j < \deg y_{j+1},\quad \lambda_j
  \in \Rset,\;   j=1,2,\ldots.
\end{equation}
Moreover, we say that a sequence of polynomials
$\{y_j\}_{j=1}^\infty$ has codimension $m$ if
\begin{equation}
  \label{eq:degyjm}
     \deg y_j = m+j-1,\quad j=1,2,\ldots
  \end{equation}
  We also say that $T(y)$ is an $m$-PES operator if the
  eigenpolynomials satisfy the above condition.  We say that $T(y)$ is
  primitive, if the eigenpolynomials do not possess a common root
  (real or complex).

Note: if $T(y)$ has at least 3 linearly independent polynomial
eigenfunctions, then necessarily, $p(x), q(x), r(x)$ must be rational
functions.

Let $I=(x_1, x_2)$ be an open interval (bounded, unbounded, or
semi-bounded) and let $W dx$ be a positive measure on $I$ with
finite moments of all orders.  We say that a sequence of real
polynomials $\{y_j\}_{j=1}^\infty$ forms an \emph{orthogonal
  polynomial system} (OPS for short) if the polynomials constitute an
orthogonal basis of the Hilbert space $\rL^2(I,W dx)$.  If
\eref{eq:degyjm} holds, we speak of an $m$-OPS.

The following definition encapsulates the notion of a system of
orthogonal polynomials defined by a second-order differential
equation.  Consider a boundary value problem
\begin{eqnarray}
  \label{eq:bvp}
  -(Py')'+ Ry  = \lambda W y\\
  \label{eq:pslpbc}
  \lim_{x\to x_{i}^{\pm}} (Py'u-Pu'y)(x) = 0,\quad i=1,2,
\end{eqnarray}
where $P(x), W(x)>0$ on the interval $I=(x_1,x_2)$, and where $u(x)$
is a fixed polynomial solution of \eref{eq:bvp}.  We speak of a
polynomial Sturm-Liouville problem (PSLP) if the resulting spectral
problem is self-adjoint, pure-point and if all eigenfunctions are
polynomial.  We speak of an $m$-PSLP if the eigenpolynomials satisfy
\eref{eq:degyjm}.  If $m=0$, then we recover the classical orthogonal
polynomials, the totality of which is delineated by Bochner's theorem.
For $m>0$, Bochner's theorem no longer applies and we encounter a
generalized class of polynomials; we name these exceptional, or $X_m$
polynomials.

Given a PSLP, the operator
\[ T(y) = W^{-1}(Py')' - W^{-1} Ry \] is PES.  Letting $p(x), q(x),
r(x)$ be the rational coefficients of $T(y)$ as in
\eref{eq:Tdef}, we have
\begin{eqnarray}
  \label{eq:Pdef}
  P(x) = \exp\left(\int^x  \!\!q/p\right),\\
  \label{eq:Wdef}
  W(x) = (P/p)(x),\\
  \label{eq:Rdef}
  R(x) = -(rW)(x),
\end{eqnarray}
Hence, for a PSLP, $P(x), R(x), W(x)$ belong to the quasi-rational
class\cite{gibbonsveselov}, meaning that their logarithmic derivative
is a rational function.

Conversely, given a PES operator $T(y)$ and an interval $I=(x_1,x_2)$
we formulate a PLSP \eref{eq:bvp} by employing
\eref{eq:Pdef}--\eref{eq:Rdef} as definitions, and by adjoining the
following assumptions:
\begin{enumerate}
 \item $P(x), W(x)$ are continuous and positive on $I$
 \item $Wdx$ has finite moments, i.e. $\int_I x^n W(x) dx <\infty,\quad  n=0,1,2,\dots$
 \item  $ \lim_{x\to x_{i}} P(x) x^n = 0,\quad i=1,2,\quad n=0,1,2\ldots$
\item the eigenpolynomials of $T(y)$ are dense in the Hilbert space $\rL^2(I,Wdx)$.
\end{enumerate}
 These definitions and assumptions (i) and
(ii) imply Green's formula:
\begin{equation}
  \label{eq:Msymmetric}
  \int^{x_2}_{x_1}  T(f)g \, Wdx - \int_{x_1}^{x_1} T(g) f\, W dx=  P(f'g-fg') \Big|^{x_2}_{x_1}
\end{equation}
By (iii) if $f(x),g(x)$ are polynomials, then the right-hand side is
zero.  If $f$ and $g$ are \emph{eigenpolynomials} of $T(y)$ with
unequal eigenvalues, then necessarily, they are orthogonal in
$\rL^2(I, Wdx)$.  Finally, by (iv) the eigenpolynomials of $T(y)$ are
complete in $\rL^2(I,Wdx)$, and hence satisfy the definition of an
OPS.

We now describe a construction that systematically generates
polynomial Sturm-Liouville systems of arbitrarily large codimension
$m\geq 0$.
\section{The Darboux transformation}
Let $T(y)$ be a differential operator \eref{eq:Tdef} with rational
coefficients.   We speak of a rational factorization if
\begin{equation}
  \label{eq:TABfactorization}
   T- \lambda_0 = B A 
\end{equation}
where  $A(y), B(y)$ are first order operators with rational
coefficients and where $\lambda_0$ is a constant.  Let us write
\begin{eqnarray}
     \label{eq:Aydef}
   A(y) = b(y'-wy),\\
   \label{eq:Bydef}
   B(y) = \hb(y'-\hw y),
\end{eqnarray}
where $w(x), \hw(x), b(x), \hb(x)$ are all rational functions.  Given
a rational factorization
we introduce the partner operator
\begin{equation}
  \label{eq:hTdef}
   \hT= A B  +\lambda_0 
\end{equation}
whose explicit form is
\begin{equation}
   \hT(y) = p y'' + \hq y' + \hr y
\end{equation}
where
\begin{eqnarray}
   \label{eq:hqdef}
   \hq = q + p' -2p b'/b\\
   \hr = -p(\hw' + \hw^2) - \hq \hw + \lambda_0 .
\end{eqnarray}
We will refer  to
\[ \phi(x) = \exp\left(\int^x\!\! w\right),\] as a factorization
eigenfunction (quasi-rational) and to $b(x)$ as the factorization
gauge (rational).  The former satisfies
\begin{equation}
  \label{eq:phi0rel}
  T(\phi) = \lambda_0 \phi.
\end{equation}
Equivalently,
\begin{equation}
  \label{eq:wxdef}
  w(x)=\phi'(x)/\phi(x)
\end{equation}
is a rational solution of the following
Ricatti-like equation:
\begin{equation}
  \label{eq:rformula}
  p (w' + w^2) + q w +r= \lambda_0.
\end{equation}

For a fixed $T(y)$, a rational factorization is fully determined by a
quasi-rational factorization eigenfunction and a rational
factorization gauge.  Indeed, given $p,q,r, w, b$ relation
\eref{eq:TABfactorization} gives us
\begin{eqnarray}
  \label{eq:hbdef}
  \hb = p/b,\\
  \label{eq:hwdef}
  \hw= -w-q/p + b'/b.
\end{eqnarray}
The choice of $b(x)$ determines the gauge of the partner operator.
Consider two factorization gauges $b_1(x), b_2(x)$ and let
$\hT_1(y), \hT_2(y)$ be the corresponding partner operators.  Then,
\[ \hT_2 = \mu^{-1} \hT_1\mu,\]
where
\[ \mu(x) = b_1(x)/ b_2(x).\]

The above construction of the partner operator is symmetric with
respect to the interchange of the hatted and unhatted variables.
Letting $P(x)$ be as in \eref{eq:Pdef}, and setting
\begin{equation}
  \label{eq:hphidef}
  \hphi(x) = \exp\left(\int^x \!\!\hw\right) =\frac{1}{(P/b)(x)\phi(x)},
\end{equation}
we  have
\[   \hT(\hphi) = \lambda_0 \hphi.\]
We also have
\begin{eqnarray}
  \label{eq:Pbrel}
  P/b = \hat{P}/\hb,\\ 
  \label{eq:bhbprel}
  \hb b = p,\\ 
  \hq/p - \hb'/\hb = q/p - b'/b
\end{eqnarray}
Thus, starting with $\hT$ and taking $\hphi$ as the factorization
function and $\hb(x)$ as the factorization gauge, we recover $T(y)$.

Next, suppose that $T$ is a PES operator with eigenpolynomials $\{ y_j
\}$.  If $\mu(x)$ is a polynomial, then $\mu T \mu^{-1}$ is also a PES
operator, with eigenpolynomials $\{\mu y_j\}$.  Therefore, we can fix the
gauge of a PES operator by requiring that the eigenpolynomials are
primitive (no common roots).    

By construction, partner operators obey the following intertwining
relations:
\begin{equation}
  \hT A = A T,\qquad
  B \hT = T B.
\end{equation}
Hence, if $T$ is a PES operator with eigenpolynomials $\{ y_j\}$, then
$\{A(y_j)\}$ are eigenfunctions of the partner operator $\hT$ with the
same eigenvalues.  By inspection of \eref{eq:Aydef}, with the
appropriate choice of $b(x)$, the $A(y_j)$ are polynomials. Hence, if
$T$ is PES, then so is $\hT$.  Furthermore, the requirement that the
eigenpolynomials of $\hT$ be primitive fixes $b(x)$ up to a choice of
scalar multiple. In many cases, such as the factorization shown in
\eref{eq:L1factorization}-\eref{eq:L1factorization2}, it will suffice
to take $b(x)$ to be the denominator of $w(x)$.  However, there are
other cases, such as the factorization shown in
\eref{eq:LfactorI-1}-\eref{eq:LfactorI-4}, where $b(x)$ must be a
rational function.

The duality between $T$ and $\hT$ has another aspect.  Let $W(x)$ be
as in \eref{eq:Wdef} and let $\hW(x)$ be analogously defined.  Hence, by
equations \eref{eq:Pbrel} \eref{eq:bhbprel}, 
\begin{equation}
  \label{eq:hWrel}
  \hW = P/b^2 = pW/b^2
\end{equation}
Consequently, $A$ and
$-B$ are formally adjoint relative to these measures:
\begin{equation}
  \label{eq:ABadjformal}
  \int^{x_2}_{x_1}  A(f) g\,\hW dx +
  \int^{x_2}_{x_1}  B(g) f\, Wdx  =  (P/b) f g \Big|^{x_2}_{x_1}
\end{equation}
If the above RHS vanishes for polynomial $f,g$, then $A$ and $-B$,
with suitably defined domains, give rise to adjoint operators in the
rigorous sense of densely defined linear operators on Hilbert spaces
$\rL^2(I,Wdx)$ and $\rL^2(I,\hW dx)$, respectively.   

Darboux transformations can be classified into three types as far as
their spectral properties are concerned \cite{deift,sukumar}:
state-deleting, state-adding, or isospectral.
\begin{enumerate}
\item \textbf{state-deleting transformation:} In this case the
  factorizing function $\phi(x)$ satisfies $\phi\in L^2(I,W dx)$ and
  the formal factorizing eigenvalue $\lambda_0$ is the
  maximum\footnote{Note that, as opposed to the usual convention in
    Schr\"odinger operators where the spectrum is bounded from below,
    in this paper the spectrum of all Sturm-Liouville problems is
    bounded from above. The eigenfunction corresponding to the maximum
    of the spectrum corresponds therefore to the \textit{ground
      state}.} of the spectrum of $T$.
\item \textbf{state-adding transformation:} In this case the partner
  factorizing function $\hat\phi(x)$ satisfies $\hat\phi\in L^2(I,\hW
  dx)$ and the formal factorizing eigenvalue $\lambda_0$ must be above
  the maximum of the spectrum of $T$. Equivalently, from
  \eref{eq:hphidef} and \eref{eq:hWrel} it follows that
  \[ \hat\phi\in L^2(I,\hW dx) \, \Leftrightarrow \, \frac{p^{1/2}}{P}
  \phi^{-1}\in L^2(I,W dx) \] so it is clear that the spectral
  properties of the transformation only depend on the choice of
  $\phi$, not on the choice of gauge $b(x)$.
\item \textbf{isospectral transformation:} In this case $\phi\notin
  L^2(I,W dx)$, $\hat\phi\notin L^2(I,\hat W dx)$ and the formal
  factorizing eigenvalue $\lambda_0$ must be above the maximum of the
  spectrum of $T$.
\end{enumerate}

In the context of algebraic Darboux transformations discussed in this
paper, if we assume that both $T$ and $\hat T$ are PSLPs, the above
spectral characterization can be particularized to a purely algebraic
one.
\begin{enumerate}
\item A state-deleting transformation corresponds to $\phi=y_1$, the
  first eigenpolynomial of $T$.
\item  A state-adding transformation corresponds to $\hat\phi$ (as defined by \eref{eq:hphidef}) being a polynomial.
\item Isospectral transformations correspond to neither $\phi$ nor
  $\hat\phi$ being polynomials.
\end{enumerate}

The above conditions can be explicitly verified on the factorizations
performed in Sections 4 and 5.  For example, equations
\eref{eq:L1factorization}-\eref{eq:L1factorization2} show an
isospectral factorization; neither of the factorizing eigenfunctions
is a polynomial.  By contrast, equations \eref{eq:LfactorI-1}-\eref{eq:LfactorI-4} show a state-deleting/state-adding
factorization; one of the factorizing eigenfunctions is a polynomial, while its
partner eigenfunction is not.

State-adding and state-deleting factorizations are
dual notions, in the sense that if the factorization of $T$ is
state-deleting, then the factorization of $\hT$ is state-adding, and
vice versa.  

As we already pointed out, the eigenpolynomials $\{
y_j \}$ and $\{\hy_j\}$ constitute orthogonal polynomial systems
relative to $\rL^2(I,Wdx)$ and $\rL^2(I,\hW dx)$, respectively.  The
adjoint relation between $A$ and $B$ allows us to compare the $\rL^2$
norms of the two families.  Indeed, by \eref{eq:TABfactorization}
\eref{eq:Tyjeigenfunctions} \eref{eq:ABadjformal},
\begin{eqnarray}
  \label{eq:Ayjnorm}
  \int_I (A(y_j))^2 \, \hW dx  = - \int_I  B(A(y_j)) y_j\,  W dx =
  (\lambda_0-\lambda_j) \int_I  y_j^2 \, Wdx  
\end{eqnarray}

\subsection{Shape-invariance}
Suppose that
\begin{equation}
  \label{eq:Tkydef}
   T_k(y) = p(x) y'' + q_k(x) y' + r_k(x) y,\quad k\in K,
\end{equation}
is a family of PES operators, where $K$ is some parameter index set.
If this family is closed with respect to the state-deleting Darboux
transformation, we speak of \emph{shape-invariant} operators and
polynomials.  To be more precise, let $\pi_k(x)= y_{k,1}(x)$ be be the
corresponding ground-state eigenpolynomial. Without loss of
generality, we assume that that the ground-state energy is zero.  and
let
\begin{equation}
  \label{eq:Tkfact}
  T_k = B_k A_k, \qquad A_k(\pi_k) = 0
\end{equation}
be the corresponding factorization.  Shape-invariance means that there
exists a one-to-one map $h:K\to K$ and real constants $\lambda_k$ such that
\begin{equation}
  \label{eq:Thkfact}
 T_{h(k)}= A_k B_k +\lambda_k.
\end{equation}
Necessarily, there exist constants $\alpha_{k,j}, \beta_{k,j}$ such that
\begin{eqnarray}
  y_{h(k),j-1} = \alpha_{k,j} A_k(y_{k,j}),\quad j\geq 1,\\
  y_{k,j+1} =\beta_{k,j+1} B_k(y_{h(k),j}),\quad j\geq 0,\\
  \beta_{k,j} \alpha_{k,j}  = \lambda_{k,j}\\
  \lambda_{h(k),j}=  \lambda_{k,j+1}  + \lambda_k.
\end{eqnarray}

In accordance with \eref{eq:Pdef}, define
\begin{equation}
  \label{eq:Pkdef}
  P_k(x) = \exp\left(\int^x \!\!q_k/p\right)
\end{equation}
Let $b_k(x)$ denote the shape-invariant factorization gauge; i.e.;
\begin{equation}
  \label{eq:Akdef}
  A_k(y) = (b_k/\pi_k) \cW(\pi_k, y),
\end{equation}
where
\begin{equation}
  \cW(f,g) = f g' - f' g
\end{equation}
Equation \eref{eq:hqdef} implies the following necessary condition,
\begin{equation}
  \label{eq:bk2rel}
  p\,P_k/P_{h(k)}  = b_k^2.
\end{equation}
This is a rather strong constraint, because the left-hand side is a
product of quasi-rational functions, while the right-hand side is a
rational squared.

\subsection{Covariant factorization}
Next, we introduce the notion of a covariant isospectral factorization.
Let $T_k(y)$ be a shape-invariant family of PES operators, as above.
Suppose that $\phi_k(x)$ is an indexed family of isospectral
factorization functions.  Let
\begin{equation}
  \label{eq:Tkisofact}
  T_k = \tB_k \tA_k + \tlambda_k,\quad     \tA_k(\phi_k) = 0
\end{equation}
be the corresponding isospectral factorization.  Let
\begin{equation}
  \label{eq:hTkisofact}
  \hT_k = \tA_k \tB_k + \tlambda_k,\quad \tB_k(\hphi_k) = 0
\end{equation}
be the partner operator and partner eigenfunction, respectively.  We
say that the factorization with respect to $\phi_k$ is covariant
if
\begin{equation}
  \label{eq:phishapeinv}
  A_k(\phi_k) \propto \phi_{h(k)}.
\end{equation}
The following Lemma furnishes a useful test for covariant
factorization.  Let us say that a PES operator is formally
non-degenerate if for every formal eigenvalue (eigenfunction is
quasi-rational) $\lambda_0\in \Rset$ there exists at most one linearly
independent quasi-rational eigenfunction with that eigenvalue.
\begin{lem}
  \label{lem:shapeinv}
  Suppose that $\phi_k(x)$ is continuous with respect to $k$ and
  formally non-degenerate for generic values of $k\in K$.
  Furthermore, suppose that 
  \begin{equation}
    \label{eq:tlambdashapeinv}
    \tlambda_{h(k)} = \tlambda_k + \lambda_k.
  \end{equation}
  Then, the factorization with respect to $\phi_k$ is covariant.
\end{lem}
\emph{Proof.}
By \eref{eq:Tkfact} \eref{eq:Thkfact}  and \eref{eq:tlambdashapeinv},
\begin{eqnarray}
  T_{h(k)} (A_k(\phi_k)) &= A_k (T_k(\phi_k)) +\lambda_k A_k(\phi_k)\\
  &=(\lambda_k+\tlambda_k) A_k(\phi_k)\\
  &= \tlambda_{h(k)} A_k(\phi_k)
\end{eqnarray}
Since $\phi_k$ is quasi-rational and since $A_k$ has rational
coefficients, $A_k(\phi_k)$ is also quasi-rational.  Hence,
\eref{eq:phishapeinv} holds for
generic $k$. Therefore, it holds for all $k$.  QED

\section{Laguerre polynomials}
Let us introduce the PES operator
\begin{equation}
  \label{eq:Lkydef}
    \cL_{k}(y):= xy''+(k+1-x)y'.
\end{equation}
The classical associated Laguerre polynomials, $\Lag{n}{k}(x)$ can be
defined as the corresponding eigenpolynomials,
\begin{equation}
  \label{eq:Lnkeigenvalue}
   \cL_k(\Lag{n}{k}) = -n \Lag{n}{k},
\end{equation}
normalized by the condition
\[ \Lag{n}{k}(x) = \frac{(-1)^n}{n!} x^n + \mbox{lower degree
  terms.} \]
The classical Laguerre polynomials are shape-invariant by virtue of
the following factorizations
\begin{eqnarray}
  \label{eq:Lkfact}
  \cL_k = B_k A_k \\
  \cL_{k+1} = A_k B_k + 1\qquad \mbox{where}\\
  \label{eq:Akydef}
  A_k(y) = y'\\
  \label{eq:Bkydef}
  B_k(y) = x y' +(k+1-x) y
\end{eqnarray}
For $k >-1$, the resulting polynomials are orthogonal relative to the
weight
\begin{equation}\label{eq:Wk}
 W_k(x)=x^k\,{\rm e}^{-x},\quad x\in (0,\infty),
\end{equation}
and can be realized as solutions of a spectral problem
\cite{atkinson,EKLW}.  The corresponding $\rL^2$ norms are given by
\begin{equation}
  \label{eq:Lnknorm}
  \int_0^\infty \Lag{n}{k}^2 W_k \,dx= \Gamma(n+k+1)/n!.
\end{equation}

The quasi-rational eigenfunctions of $\cL_k(y)$ are known
\cite[Sec. 6.1]{bateman}:
\begin{eqnarray}
  \label{eq:chgpolys1}
    \phi_1(x)&= \Lag{m}{k}(x), & \lambda_0 = -m\\
    \phi_2(x)&=x^{-k} \Lag{m}{-k}(x) & \lambda_0 =k-m\\
    \phi_3(x)&= e^x \Lag{m}{k}(-x) & \lambda_0 = k+1+m\\
  \label{eq:chgpolys4}
    \phi_4(x)&= x^{-k}e^x \Lag{m}{-k}(-x),\quad & \lambda_0 = m+1,
\end{eqnarray}
where $m=0,1,2,\ldots$.  The corresponding factorizations were
analyzed in \cite{GKMDarboux1}. Of these, $\phi_1$ with $m=0$
corresponds to a state-deleting transformation and underlies the
shape-invariance of the classical Laguerre polynomials.  For $m>0$,
the $\phi_1$ eigenfunctions yield singular operators and hence do not
yield novel orthogonal polynomials.  The $\phi_4$ family results in a
state-adding transformation.  The resulting orthogonal polynomials do
not satisfy condition \eref{eq:degyjm}; such factorizations were
discussed in \cite{GKMDarboux2}.  The type 2 and type 3 factorizations
$\phi_2, \phi_3$ result in novel orthogonal polynomials, although for
$\phi_3$ it is necessary to assume that $k>m$.  These families
correspond, respectively, to the type L1, L2 Laguerre polynomials of
\cite{sasaki3}.

Let us consider these two families of factorization on a case-by-case
basis.  The derivations that follow depend in an elementary fashion on
the following well-known identities of the Laguerre polynomials.  We
will apply them below without further comment.
\begin{eqnarray}
  \Lag{0}{k}(x) = 1\\
  \label{eq:Lneg}
  \Lag{n}{k}(x) = 0,\quad n\leq -1,\\
  \label{eq:id3}  
  n\Lag{n}{k}(x)+(x-2n-k+1)\Lag{n-1}{k}(x)+(n+k-1)\Lag{n-2}{k}(x)=0,\\
  \Lag{n}{k}'(x)=-\Lag{n-1}{k+1}(x), \label{eq:id1}\\
  \Lag{n}{k}(x)=\Lag{n}{k+1}(x)-\Lag{n-1}{k+1}(x). \label{eq:id2}
\end{eqnarray}

\subsection{The L1 family}
Fix an integer $m\geq 0$ and a real $k>-1$.
Take $\phi_3(x)$ as the factorization function and
\begin{equation}
  \label{eq:xikm}
  \xi_{k,m}(x)=\Lag{m}{k}(-x)
\end{equation}
as the factorization gauge.  Applying \eref{eq:Aydef}
\eref{eq:Bydef} \eref{eq:hbdef} \eref{eq:hwdef}, the resulting
factorization is
\begin{eqnarray}
  \label{eq:L1factorization}
  \cL_{k} = B^{\rI}_{k,m}A^{\rI}_{k,m} +k+m+1 ,\quad\mbox{where}\\
  A^{\rI}_{k,m}(y) = \xi_{k,m} y' - \xi_{k+1,m} y  \label{eq:L1factorization1}\\
  B^{\rI}_{k,m}(y) = (x y' + (1+k) y)/\xi_{k,m} \label{eq:L1factorization2}
\end{eqnarray}
The partner eigenfunction is $\hphi(z) = z^{-1-k}$. Let us define
\begin{eqnarray}
 \label{eq:Lkmdef}
 \cLI{k,m} = A^{\rI}_{k-1,m}B^{\rI}_{k-1,m} + k+m \\
 \cLI{k,m}(y) = xy''+(k+1-x)y' + my - 2\rho_{k-1,m} (x y' + k y),\quad
 \mbox{where}\\ 
 \rho_{k,m} = \xi_{k,m}'/\xi_{k,m} = \xi_{k+1,m-1}/\xi_{k,m}.
\end{eqnarray}
On the basis of the above factorization, we define type I exceptional
Laguerre polynomials to be
\begin{eqnarray}\label{eq:defXI}
  \XLagI{n}{k}{m}&=-A^{\rI}_{k-1,m}\left(\Lag{n-m}{k-1}\right)\\
  &=\xi_{k,m}\,\Lag{n-m}{k-1} + \xi_{k-1,m}\,\Lag{n-m-1}{k},\qquad
  n\geq m
\end{eqnarray}
By construction, these polynomials satisfy
\begin{equation}
  \cLI{k,m}(\XLagI{n}{k}{m}) =(m-n) \XLagI{n}{k}{m},\quad n\geq m.
\end{equation}
By \eref{eq:Msymmetric} and \eref{eq:hWrel} the sequence $\{
\XLagI{n}{k}{m}\}_{n=m}^\infty$ constitutes an $m$-OPS relative to the
weight
\begin{equation}
  \label{eq:WIdef}
  W^{\rI}_{k,m}(x) = x^k e^{-x}/ \xi_{k-1,m}^2,\quad x\in (0,\infty)
\end{equation}
Using \eref{eq:Ayjnorm} and \eref{eq:Lnknorm}, we obtain
\begin{equation}
  \label{eq:LInorm}
  \int_0^\infty (\XLagI{n}{k}{m})^2\, W^{\rI}_{k,m}\, dx = (k+n)
  \Gamma(k+n-m)/(n-m)! 
\end{equation}
For $m=0$ the above definitions reduce to their
classical counterparts; to wit,
\begin{eqnarray}
    \cLI{k,0} = \cL_k\\
    \XLagI{n}{k}{0} = \Lag{n}{k},\\
    W^{\rI}_{k,0}(x) = x^k e^x
\end{eqnarray}

\subsection{The L2 family} Fix an integer $m\geq 0$ and $k> m$, and
take $\phi_2(x)$ as the factorization function. Set
\begin{equation}
  \label{eq:etakmdef}
  \eta_{k,m}(x)=\Lag{m}{-k}(x);
\end{equation} 
take $x\, \eta_{k,m}$ as the factorization gauge. The
resulting factorization is
\begin{eqnarray}
  \label{eq:L2factorization}
  \cL_{k} = B^{\rII}_{k,m}A^{\rII}_{k,m} +(k-m) ,\quad
  \mbox{where}\\
  A^{\rII}_{k,m}(y) = x \eta_{k,m} y' +(k-m) \eta_{k+1,m} y\\
  B^{\rII}_{k,m}(y) = ( y' -y)/\eta_{k,m}
\end{eqnarray}
The partner eigenfunction is $\hphi(z) = e^z$. Based on this
factorization, we define
\begin{eqnarray}
 \label{eq:Lkmdef}
 \cL^{\rII}_{k,m} = A^{\rII}_{k+1,m}B^{\rII}_{k+1,m} + (k+1-m) \\ 
 \cL^{\rII}_{k,m}(y) = xy''+(k+1-x)y' -m y + 2x\sigma_{k+1,m} (y'-y),\quad \mbox{where}\\
 \sigma_{k,m}= -\eta_{k,m}' / \eta_{k,m} = \eta_{k-1,m-1}/\eta_{k,m}
\end{eqnarray}
We now define the type II $X_m$
Laguerre polynomials to be
\begin{eqnarray}\label{eq:defXI}
  \XLagII{n}{k}{m}&=-A^{\rII}_{k+1,m}\left(\Lag{n-m}{k+1}\right),\qquad
  n\geq m\\
  &=x\eta_{k+1,m}\,\Lag{n-m-1}{k+2} + (m-k-1)\eta_{k+2,m}\,\Lag{n-m}{k+1}
\end{eqnarray}
By construction, these polynomial satisfy
\begin{equation}
  \cLII{k,m}(\XLagI{n}{k}{m}) =(m-n) \XLagI{n}{k}{m},\quad n\geq m.
\end{equation}
Thus, the sequence $\{ \XLagII{n}{k}{m}\}_{n=m}^\infty$ constitutes an
$m$-OPS relative to the weight
\begin{equation}
  \label{eq:WIIdef}
  W^{\rII}_{k,m}(x) = x^k e^{-x}/ \eta_{k+1,m}^2,\quad x\in (0,\infty)
\end{equation}
Using \eref{eq:Lnknorm}, we also have
\begin{equation}
  \label{eq:LIInorm}
  \int_0^\infty (\XLagII{n}{k}{m})^2\, W^{\rII}_{k,m}\, dx = \frac{(1+k+n-2m)}{(n-m)!}\,
  \Gamma(2+k+n-m) 
\end{equation}
As above, for $m=0$ the above definitions reduce to their classical
counterparts, albeit the polynomials have a different normalization:
\begin{equation}
    \XLagII{n}{k}{0} = -(k+1+n) \Lag{n}{k}.
\end{equation}
The proof that the sets $\{\XLagI{n}{k}{m}\}_{n=m}^\infty$ and
$\{\XLagII{n}{k}{m}\}_{n=m}^\infty$ span dense subspaces of the
Hilbert spaces $L^2([0,\infty),W^{\rI}_{k,m}dx)$ and
$L^2([0,\infty),W^{\rII}_{k,m}dx)$ will be given in a forthcoming
publication \cite{GKM10}.

\section{Shape-invariance of the exceptional polynomials}

In this section we prove that above defined $X_m$ polynomials are
shape-invariant.  The explanation for this remarkable fact is the
commutativity/permutability of iterated Darboux transformations, also
known as the Darboux-Crum transformation.

Let $T_0(y) = T(y)$ be a given PES operator, and let
$\phi_1(x),\ldots, \phi_n(x)$ be quasi-rational eigenfunctions.  Let
$T_1(y) = \hT(y)$ be the partner PES operator corresponding to
$\phi_1$.  Now $A_1(\phi_2)$ is a quasi-rational eigenfunction for
$T_1(y)$.  Let $T_2(y) = \hT_1(y)$ be the corresponding partner
operator.  Continue in like fashion.  We arrive at the following chain
of factorizations:
\begin{eqnarray}
  T_0 &= B_1A_1 + \lambda_1\\
  T_j &= A_jB_j + \lambda_j,\quad j=1,\ldots, n-1,\\
  &= B_{j+1} A_{j+1} + \lambda_{j+1}\\
  T_n &= A_n B_n + \lambda_n,
\end{eqnarray}
where
\begin{equation}
  (A_j\cdots A_2 A_1)(\phi_j) = 0,\quad j=1,2,\ldots,n.
\end{equation}
In the end, we obtain the following intertwining relations:
\begin{eqnarray}
  T_0\cB =  \cB T_n,\qquad\mbox{where }\quad \cB =  B_1 B_2 \cdots B_n\\
  \cA T_0 = T_n \cA,\qquad \mbox{where }\quad \cA=A_n \cdots A_2 A_1.
\end{eqnarray}
By construction, 
\begin{equation}
  \cA(\phi_j) = 0,\quad j=1,2,\ldots, n.
\end{equation}
Hence,
\begin{equation}
  \cA(y) = b(x) \cW(\phi_1,\ldots, \phi_n,y)/\cW(\phi_1,\ldots, \phi_n),
\end{equation}
where $b(x)$ is the higher-order rational factorization gauge, and
where $\cW$ denotes the Wronskian operator.  
As before, $b(x)$ is uniquely
determined (up to scalar multiple) by the requirement that the
eigenfunctions of $T_n(y)$ constitute a primitive sequence of
polynomials. 

The key observation is that up to sign, the above definition of $T_n$
is independent of the order of the factorization functions.  Let us
exploit this commutativity to prove that the above-defined $X_m$
polynomials are shape-invariant.  To do so, requires that we consider
a certain 2-step factorization.

Let $T_k(y)$ be a family of shape-invariant PES operators as per
\eref{eq:Tkydef}. Let $\pi_k(x)=y_{k,1}(x)$ denote the corresponding
ground-state eigenpolynomials, and let \eref{eq:Tkfact}
\eref{eq:Thkfact} be the corresponding factorizations, where without
loss of generality the factorization eigenvalue is set to zero.

Next, let $\phi_{k}(x)$ be a quasi-rational eigenfunction that
corresponds to a covariant, isospectral factorization as per
\eref{eq:phishapeinv}.  Let $\hT_k(y)$ be the corresponding family of
isospectral operators as per \eref{eq:hTkisofact}. We claim that this family
is also shape-invariant.  Let
\begin{equation}
  \hT_k = \hB_k \hA_k,\qquad \hA_k(\tpi_k) = 0
\end{equation}
be the ground-state factorization of the partner operator, where
\begin{equation}
  \tpi_k = \tA_k(\pi_k) 
\end{equation}
is the new ground-state polynomial.  Our claim is that
\begin{equation}
  \label{eq:hTshapeinv}
    \hT_{h(k)} = \hA_k \hB_k+\lambda_k,
\end{equation}
For convenience, let us set
\begin{equation}
  \tT_k= \hA_k \hB_k 
\end{equation}
The 2nd order intertwining relation is
\begin{eqnarray}
    \label{eq:2stepATTA}
  \cA_k T_k = \tT_k \cA_k\quad\mbox{where}\\
  \cA_k(y) = b(x) W(\pi_k ,\phi_k,y) /W(\pi_k,\phi_k),
\end{eqnarray}
and where $b(x)$ is the 2nd order rational factorization gauge whose
form is not relevant to our argument. There are two ways to factorize
$\cA_k$, the 2nd order intertwiner:
\begin{eqnarray}
  \cA_k =  \hA_k \tA_k\\
  \cA_k = \tA_{h(k)}A_k 
\end{eqnarray}
The 2nd equation is true because by \eref{eq:phishapeinv} we have
\begin{equation}
   \tA_{h(k)} (A_k(\phi_k)) \propto \tA_{h(k)} (\phi_{h(k)})= 0.
\end{equation}
Hence,
\begin{eqnarray}
  \cA_k T_k &= \tA_{h(k)} A_k T_k \\
  &= \tA_{h(k)} (T_{h(k)}-\lambda_k) A_k \\
  &= (\hT_{h(k)}-\lambda_k)  \cA_k.
\end{eqnarray}
Hence, by \Eref{eq:2stepATTA},
\begin{equation}
  \tT_k \cA_k =   (\hT_{h(k)}-\lambda_k)  \cA_k.
\end{equation}
The ring of differential operators with rational coefficients has no
zero divisors. Therefore, the desired relation \eref{eq:hTshapeinv}
follows.

Next, let us illustrate the above result by explicitly showing the
shape-invariant factorization for the type I exceptional Laguerre
polynomials defined in the preceding section.  The index set consists
of real $k> -1$. Let us set
\begin{eqnarray}
  \pi_k(x) = 1,\\
  h(k)=k+1\\
  \lambda_k=1\\
  T_k(y) = \cL_k(y)
\end{eqnarray}
The classical Laguerre polynomials are shape-invariant; relations
\eref{eq:Tkfact} \eref{eq:Thkfact} hold, as per
\eref{eq:Lkfact}-\eref{eq:Bkydef}.
Let us fix an integer $m\geq 0$ and set
\begin{eqnarray}
  \phi_k(x) = e^x \xi_{k,m},  \label{eq:phi3}\\
  \tA_k(y) = A^{\rI}_{k,m}(y)\\
  \tB_k(y) = B^{\rI}_{k,m}(y)\\
  \hT_k(y) = \cL^{\rI}_{k+1,m}\\
  \tlambda_k = k+m
\end{eqnarray}
These definitions realize a particular instance of the isospectral
factorizations shown in \eref{eq:Tkisofact} \eref{eq:hTkisofact}.  By
inspection of \eref{eq:chgpolys1}-\eref{eq:chgpolys4}, the operator
$\cL_k$ is formally non-degenerate for generic $k$.  Equation
\eref{eq:tlambdashapeinv} is satisfied, and hence by Lemma
\ref{lem:shapeinv} the isospectral factorization with respect to
\eref{eq:phi3} is covariant.  Therefore, the operators
$\cL^{\rI}_{k,m}$ are shape-invariant.

Next, we explicitly describe the ground-state factorization for
$\cL^{\rI}_{k,m}$ and verify the shape-invariance property.  To
determine an explicit form for $\hA_k , \hB_k$ we make use of formula
\eref{eq:bk2rel}. Here,
\begin{eqnarray}
  \hq_{k}(x) = (k+2-x) -2 x \rho_{k+1,m} \\
  \hP_{k}(x) = \exp\left(\int^x \!\!\hq_k/p\right) = e^{-x} x^{k+2}/\xi_{k,m}^2\\
  p \,\hP_k/\hP_{k+1} = \xi_{k+1,m}^2/\xi_{k,m}^2
\end{eqnarray}
In this way, we arrive at the shape-invariant factorization
\begin{eqnarray}
  \cL^{\rI}_{k,m} = \hB^{\rI}_{k,m} \hA^{\rI}_{k,m},\quad\label{eq:LfactorI-1}
  \hA^{\rI}_{k,m}(\xi_{k,m}) = 0,\\  
  \label{eq:LfactorI-2} 
  \cL^{\rI}_{k+1,m} = \hA_{k,m} \hB_{k,m}+1,  \quad
  \hB^{\rI}_{k,m}(e^x x^{-1-k} \xi_{k-1,m}) = 0  
\end{eqnarray}
where
\begin{eqnarray}
  \hA^{\rI}_{k,m}(y) = (\xi_{k,m}/\xi_{k-1,m}) (y'-\rho_{k,m}\, y)\\
  \hB^{\rI}_{k,m}(y) =(\xi_{k-1,m}/\xi_{k,m}) (xy'+(1+k)\, y)- x
  y\label{eq:LfactorI-4} 
\end{eqnarray}
Thus, the type I polynomials obey the following lowering and raising relations:
\begin{eqnarray}
  \label{eq:LIraiselower}
  \hA^{\rI}_{k,m}(\XLagI{n}{k}{m}) = - \XLagI{n-1}{k+1}{m},\quad n\geq m.\\
  \hB^{\rI}_{k,m}(\XLagI{n}{k+1}{m}) = (n+1-m)
  \XLagI{n+1}{k}{m},\quad n\geq m.
\end{eqnarray}

In a similar fashion, we derive the following shape-invariant
factorization for the type II polynomials.  This time, we let
\begin{eqnarray}
  \phi_k(x) = x^{-k} \eta_{k,m},\\
  \tlambda_k = k-m,\\
  \hT_k(y) = \cL^{\rII}_{k-1,m},\\
  \tA_k(y) = A^{\rII}_{k,m},\\
  \tB_k(y) = B^{\rII}_{k,m},\\
  \hq_{k}(x) = (k-x) +2 x\sigma_{k,m} \\
  \hP_{k}(x) = \exp\left(\int^x \!\!\hq_k/p\right) =  e^{-x} x^{k}/\eta_{k,m}^2\\
  p\, \hP_k/\hP_{k+1} = \eta_{k+1,m}^2/\eta_{k,m}^2
\end{eqnarray}
Applying the formulas of section 3, we obtain the following
shape-invariant factorization:
\begin{eqnarray}
  \cL^{\rII}_{k,m} = \hB^{\rII}_{k,m} \hA^{\rII}_{k,m},\qquad
  \hA^{\rII}_{k,m}(\eta_{k+2,m}) = 0,\\  
  \cL^{\rII}_{k+1,m} = \hA^{\rII}_{k,m} \hB^{\rII}_{k,m}+1,  \quad
  \hB^{\rII}_{k,m}(e^x x^{-1-k} \eta_{k+1,m}) = 0
\end{eqnarray}
where
\begin{eqnarray}
  \hA^{\rII}_{k,m}(y) = (\eta_{k+2,m}/\eta_{k+1,m}) (y'+\sigma_{k+2,m}\, y)\\
  \hB^{\rII}_{k,m}(y) =(\eta_{k+1,m}/\eta_{k+2,m}) (xy'+(1+k-x)\, y)+\\
  \qquad\qquad +(\eta_{k,m-1}/\eta_{k+2,m})  x y
\end{eqnarray}
The type II polynomials obey the same lowering and raising relations
as in \eref{eq:LIraiselower}.

\ack The research of DGU was supported in part by MICINN-FEDER grant
MTM2009-06973 and CUR-DIUE grant 2009SGR859.  The research of NK was
supported in part by NSERC grant RGPIN 105490-2004. The research of RM
was supported in part by NSERC grant RGPIN-228057-2004.

\end{document}